\documentclass[useAMS,usenatbib,a4paper]{mn2e_mod}
\usepackage{epsfig,amsmath,amssymb,amsfonts,natbib,color,url}

\def\s{{\rm\,s}}

\def\cm{{\rm\,cm}}   
\def\m{{\rm\,m}}   
\def\km{{\rm\,km}}   
   
\def\m{{\rm\,m}}   
\def\yr{{\rm\,yr}}   
   
\def\g{{\rm\,g}}

\title[Stochastic Flights of Propellers]{Stochastic Flights of Propellers}
\author[M.~Pan,\,H.~Rein,\,E.\,Chiang,\,\&\,S.N.\,Evans]{Margaret~Pan$^{1}$\footnotemark[1],\,Hanno~Rein$^{2}$,\,Eugene~Chiang$^{1}$,\,and\,Steven~N.~Evans$^{3}$\\
 \\
  $^1$Department of Astronomy, University of California at Berkeley, Hearst Field Annex B-20, Berkeley CA 94720-3411, USA\\
  $^2$Institute for Advanced Study, 1 Einstein Drive, Princeton, NJ 08540, USA\\
  $^3$Department of Statistics, University of California at Berkeley, 367 Evans Hall, Berkeley CA 94720-3860, USA\\}
\date{Submitted: \today}

\begin{document}
\maketitle
\begin{abstract}
  Kilometer-sized moonlets in Saturn's A ring create S-shaped wakes
  called ``propellers'' in surrounding material. The {\it Cassini}
  spacecraft has tracked the motions of propellers for several years
  and finds that they deviate from Keplerian orbits with constant semimajor
  axes. The inferred orbital migration is known to switch sign.
  We show using a statistical test that the time series of orbital
  longitudes of the propeller Bl\'eriot is consistent with that of a
  time-integrated Gaussian random walk. That is, Bl\'eriot's observed
  migration pattern is consistent with being stochastic.  We further
  show, using a combination of analytic estimates and collisional
  $N$-body simulations, that stochastic migration of the right
  magnitude to explain the {\it Cassini} observations can be driven by
  encounters with ring particles 10--20 m in radius. That the local
  ring mass is concentrated in decameter-sized particles is supported on
  independent grounds by occultation analyses.

\end{abstract}

\label{firstpage}
\footnotetext[1]{e-mail: \rm{\url{mpan@astro.berkeley.edu}.}}

\section{INTRODUCTION}  
\label{sec:intro}  
  
Propellers are disturbances in Saturn's rings caused by moonlets
0.1--1 km in radius \citep{tiscarenoetal06}. The moonlets
gravitationally repel material to either side of their orbits,
creating partial gaps that diffuse shut via inter-particle collisions
\citep{spahn00,srem02,sei05,lewis09}. Even with the {\it Cassini}
spacecraft's resolving power, the moonlets themselves are too small to
be detected directly. Their existence and sizes are inferred from the
larger S-shaped wakes resembling propellers that they leave behind, on
scales of several Hill radii (\citealt{sei05}; \citealt{tiscareno10},
hereafter T10).
  
\begin{figure*}  
\begin{center}  
\includegraphics[scale=.8]{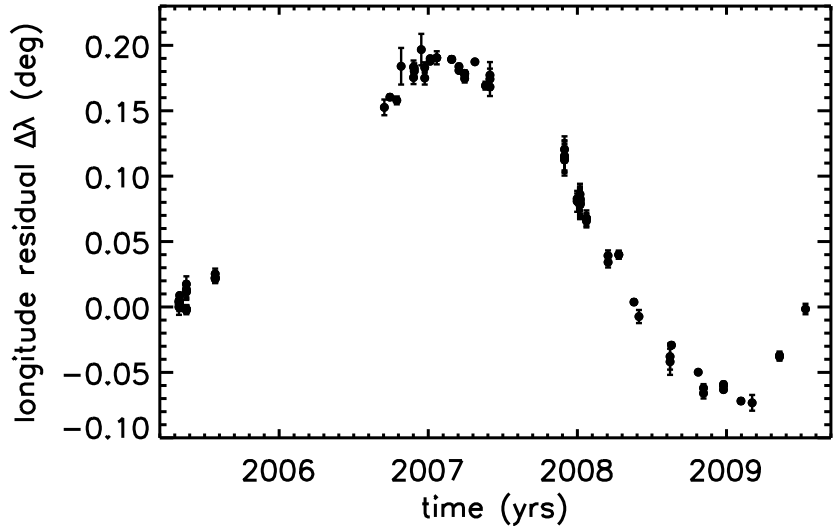} 
\end{center}  
\caption{Longitude residuals (deviations from a fixed circular orbit)
  of the propeller Bl\'{e}riot as seen in 89 {\it Cassini} images
  obtained over 4.2~years and as reported by T10. For display
  purposes, we subtracted a linear trend to place the first and last
  data points at zero longitude residual.
\label{fig:bleriot}  }
\end{figure*}

Intriguingly, multi-epoch {\it Cassini} observations of a number of  
propellers reveal that propellers deviate from  
strictly Keplerian orbits: the orbital longitudes $\lambda$ of a  
propeller drift away from the values expected for an orbit of fixed  
semimajor axis (T10).  Longitude residuals $\Delta \lambda$ range from  
0.01--0.31 degrees 
over $\Delta t = 1.3$--4.3 years; see Table 1 of T10. By far the most  
extensive data exist for the propeller dubbed Bl\'eriot, whose  
longitudes have been measured 89 times at sporadic intervals over 4.2  
years. Bl\'eriot's longitude residuals versus time are shown in  
Figure~\ref{fig:bleriot}, 
reproduced from T10. The measurements of  
non-Keplerian motion represent the first direct evidence that moons  
embedded in rings exhibit orbital evolution.  Propellers thus provide  
a test-bed for studying satellite-disk interactions, in particular  
orbital migration \citep[see, for example,][]{goldreich82}.  
 
To establish the order of magnitude of the implied migration, 
we convert longitude residual $\Delta \lambda$ to the change 
$\Delta a$ in the moonlet's semimajor axis: 
\begin{align}  
\Delta a   
& \sim \frac{\Delta \lambda}{\Omega\Delta t} a \nonumber \\  
& \sim 30 \left( \frac{\Delta \lambda}{0.1 \, {\rm deg}} \right)  
       \left( \frac{2 \, {\rm yr}}{\Delta t} \right)   
       \left( \frac{a}{1.3 \times 10^5 \km} \right)^{5/2} \, {\rm m}   
\label{eqn:delta_a}  
\end{align}  
where $\Omega$ is the mean motion (orbital frequency).  This inferred 
radial deviation should not be confused with the azimuthal deviation, 
which is measured directly from observations to be on the order of $a 
\Delta \lambda \sim 300$ km. The semimajor axis change $\Delta a$ 
observed to date is smaller than the moonlet's expected physical size. 
  
Two classes of theories have emerged to explain the non-Keplerian  
motion. In one, propeller-moonlets librate within a resonance  
established by co-orbital material \citep{pan10}. In the other,  
propeller-moonlets are torqued stochastically by density fluctuations  
in surrounding ring material \citep{rein10,crida10}.  
Both types of theories in their current forms are not without problems.  
Libration amplitudes within the proposed co-orbital resonance  
damp to zero when account is made of the two-way feedback 
between the moonlet and the ring \citep{pan12}. 
So far, the stochastic migration hypothesis has focused 
on density fluctuations driven by self-gravity (``self-gravity wakes''; 
\citealt{salo95}). \citet{rein10} simulated such wakes, 
and found that they could cause small moonlets, about 25--50 m in radius, 
to random walk by distances comparable to those cited above. 
Unfortunately, most propellers, including Bl\'eriot, are an order 
of magnitude larger in size (T10), and thus are not expected to be 
accelerated significantly by self-gravity wakes. 
This shortcoming of self-gravity wakes 
can be remedied by increasing the ring surface density (e.g., 
\citealt{crida10}), but evidence for such large surface densities is 
lacking \citep{colwell09}. 
 
It may seem surprising that even the most basic issue of whether the  
non-Keplerian motion is deterministic or random is controversial.
A glance at the longitude time series in Figure~\ref{fig:bleriot}  
suggests that Bl\'eriot's motion is smooth or even  
sinusoidal, with a period of $\sim$3.7 yr. Nevertheless, the time  
series could in fact reflect pure noise. The confusion arises  
because orbital longitude is a time-integrated quantity:  
\begin{equation}  
\Delta \lambda = \int^t \Delta \Omega \; dt  
\end{equation}  
where $\Delta \Omega$ is the difference between the moonlet's  
instantaneous mean motion and that of a fixed reference  
orbit. The time integration smooths over fluctuations in $\Delta\Omega$;  
$\Delta \lambda$ is obviously differentiable. The integration also  
introduces correlations between data points even when $\Delta\Omega$  
itself represents uncorrelated noise: $\Delta \lambda (t)$ depends on  
the full time history of perturbations up to time $t$. Both effects  
conspire to hide any underlying stochasticity.  
 
In this paper, we apply a test, well-known among statisticians but  
less so among astronomers, that identifies integrated random walks for  
the special case where the underlying random walk is Gaussian. In  
effect, the test ``undoes'' the correlations introduced by the  
integration to determine whether the integrand $\Delta\Omega$ is  
consistent with a Gaussian random walk. This test, which we call the  
``diagonalization test'' for reasons explained in \S\ref{sec:diagon},  
is especially useful because it can be applied to data that 
--- as is the case for real-life propellers 
--- are unevenly sampled.
\footnote{If the sampling were uniform, then we could in principle just
  histogram {\em second} differences of the sequence of points to
  check consistency with an integrated Gaussian random walk. Taking
  differences between adjacent points would produce the Gaussian
  random walk observed on a uniform grid of times. Taking
  differences again would produce a sequence of independent,
  identically distributed Gaussian random variables, with each one
  being the sum of some common number of the Gaussian steps in the
  underlying random walk, taken over a set of times disjoint
  from those leading to the other sums.}
  
The plan of this paper is as follows. In \S\ref{sec:diagon}, we
describe the diagonalization test and apply it to Bl\'eriot. We find
that Bl\'eriot passes the test --- that its behavior is consistent
with that of an integrated Gaussian random walk.  In
\S\ref{sec:poisson} we describe how such a random walk can be driven
by Poisson fluctuations in the encounter rate between large ring
particles and a given propeller moonlet (an effect distinct from
self-gravity wakes).  A summary is given in \S\ref{sec:sum}.
  
\section{DIAGONALIZING THE NOISE}  
\label{sec:diagon}  
  
We derive and explain the rationale behind the diagonalization test in
\S\ref{sec:decorr}, check the test against some case examples in
\S\ref{sec:test_gauss}, and apply the test to Bl\'eriot in
\S\ref{sec:bleriot}.

\subsection{Decorrelating the Integrated Gaussian Random Walk}  
 \label{sec:decorr}  
  
We wish to check if a given time series --- for example, the  
Bl\'{e}riot longitude residuals --- is consistent with a  
time-integrated random walk whose individual steps are independent  
identically-distributed (IID) Gaussian random variables. We can think  
of this time-integrated random walk as correlated Gaussian noise in  
which all the correlations arise from the time integration. If all 
such correlations were to be eliminated from our given time series, we  
could then compare what remains to a family of IID
Gaussian random variables and thereby test the  
Gaussian random walk hypothesis. Here we describe one such  
decorrelation method. The mathematical ideas behind the method are  
well known \citep[see, for example,][]{mardia79}. Indeed, the method  
overlaps conceptually with recent treatments of pulsar timing noise  
\citep{coles11}.   
  
We first calculate the covariance between any two elements  
of a time series produced by a time-integrated Gaussian random walk.  
Although our presentation focuses on the specific case of a time  
series of orbital longitudes, the underlying algorithm is general.  We  
use $\Delta$ for quantities that evolve with time and $\delta$ for  
quantities associated with an individual step in the random walk.  

Consider a body whose semimajor axis $a$ undergoes a random walk,
taking one step in the walk per timestep.  After $n\geq 1$ timesteps
of constant length $\delta t$, the total change in semimajor axis is
\begin{equation} 
\Delta a(n \delta t) = \sum_{j=0}^{n-1} \xi_j 
\end{equation} 
where the $\xi_j$ are Gaussian random variables  
for which $E[\xi_j] = 0$ and $E[\xi_j \xi_k] = (\delta a)^2 \delta_{jk}$. 
Here $E$ denotes the expected value, 
and $\delta_{jk}$ is the Kronecker delta. 
 
The longitude residual $\Delta \lambda$  
is the time integration of $\Delta a\,\partial\Omega/\partial a$: 
\begin{align}  
\Delta\lambda(n \delta t) &= -\sum_{i=1}^n \frac{3\Omega}{2a}\Delta a(i \delta t) \delta t \\ 
                 &= -\frac{3\Omega\delta t}{2a} \sum_{i=1}^n \sum_{j=0}^{i-1} \xi_j \\ 
                 &= -\frac{3\Omega\delta t}{2a} \sum_{j=0}^{n-1} (n-j) \xi_j   
\end{align}  
where, as before, $\delta t$ is the time interval between steps. 
  
The covariance between values of $\Delta \lambda$  
after $n$ and $m$ timesteps, $n<m$, is the expected value of their product:
\begin{align}  
& E[\Delta\lambda(n \delta t)\Delta\lambda(m \delta t)]\\  
& \quad = \left(\frac{3\Omega\delta t}{2a}\right)^2 
E\left[ \left(  \sum_{j=0}^{n-1} (n-j) \xi_j \right)\left(  \sum_{k=0}^{m-1} (m-k)\xi_k \right)\right]\\ 
& \quad = \left(\frac{3\Omega\delta t}{2a}\right)^2 
E\left[ \sum_{j=0}^{n-1} \xi_j \xi_j (n-j)  (m-j) \right] \label{eqn:nltm}\\ 
& \quad = \left(\frac{3\Omega\delta t \delta a}{2a}\right)^2 \, 
\sum_{j=0}^{n-1} (n-j)  (m-j) \\ 
& \quad = \left(\frac{3\Omega\delta t \delta a}{2a}\right)^2 \, 
\frac{n}{6}(1-n^2 +3m+3nm) \,. 
\label{eqn:covar}  
\end{align}  
Line (\ref{eqn:nltm}) follows because $\Delta \lambda(n \delta t)$, $\Delta  
\lambda(m \delta t)$ are two snapshots of the same integrated random walk  
$\{\Delta \lambda(\delta t),\Delta \lambda(2 \delta t),...,\Delta  
\lambda(n \delta t),...,\Delta \lambda(m \delta t),...\}$. Between timesteps $1$ and $n$,  
the histories of $\Delta \lambda(n \delta t)$ and $\Delta \lambda(m \delta t)$ coincide  
exactly --- they are 100\% correlated --- so for summation indices  
$j,k<n$, in effect $i=j$.  After timestep $n$, the Gaussian variables
$\xi_n, \ldots, \xi_{m-1}$ contributing to the further history of $\Delta  
\lambda(m \delta t)$ are completely independent of those that
contributed to $\Delta \lambda(n \delta t)$, so the determinants of the
motion after timestep $n$ do not contribute to $E[\Delta  
  \lambda(n \delta t)\Delta \lambda(m \delta t)]$.  
 
Given a list of observations at times $\{t_k : 1 \le k \le K\}$, and
choosing $\delta t$ to be the characteristic time between changes in
semimajor axis, we can use the above with $n=t_k/\delta t$ and
$m=t_\ell/\delta t$ to calculate the entries of the corresponding
(positive definite, symmetric) covariance matrix $(\Sigma_{k \ell})_{1
  \le k,\ell \le K}$.  If we write $\overrightarrow{\Delta \lambda}$
for the column vector with $k^{\mathrm{th}}$ entry $\Delta
\lambda(t_k)$, then $\Sigma$ is just the $K \times K$ matrix
$E[\overrightarrow{\Delta \lambda} \; \overrightarrow{\Delta
    \lambda}^T]$. By the spectral theorem, there is an orthogonal
matrix $U$ and a diagonal matrix $Z$ with positive diagonal entries
such that $\Sigma = UZU^T$. To eliminate the correlations in
$\Delta\lambda$ due to the time integration, we simply use this
covariance matrix to define a suitable linear transformation of the
time-series vector $\overrightarrow{\Delta\lambda}$:
\begin{align}  
\mathrm{diagonalized \; (decorrelated) \; residuals\;} \vec{r} \nonumber \\ 
= UZ^{-1/2}U^T \overrightarrow{\Delta\lambda}, 
\label{eqn:decorr}  
\end{align}  
where $Z^{-1/2}Z^{-1/2}=Z^{-1}$, the inverse of $Z$.  The covariance
matrix of such a column vector $\vec{r}$ is
\begin{align}
E[\vec{r} \; \vec{r}^T] 
& = UZ^{-1/2}U^T E[\overrightarrow{\Delta \lambda} \;  \overrightarrow{\Delta \lambda}^T] U Z^{-1/2} U^T\\
& = UZ^{-1/2}U^T \Sigma U Z^{-1/2} U^T \\
& = UZ^{-1/2}U^T U Z U^T U Z^{-1/2} U^T \\
& = I \, ,
\end{align}
the $K \times K$ identity matrix.  Because linear transformations of
Gaussian random vectors are also Gaussian (see, for example, Chapter 3 of
\citealt{mardia79}), the entries in $\vec{r}$ are IID
Gaussian random variables with common expected value $0$ and common
standard deviation $1$ --- this fact is effectively the content of
Corollary 3.2.1.1 of \cite{mardia79}.  

Note that the covariance matrix $\Sigma$ is in practice unknown, as
only the $\Delta\lambda$ are observed. However, $\Sigma$ is of the
form $\Sigma = c^2 \tilde \Sigma$, where $c = \frac{3\Omega\delta t
  \delta a}{2a}$ and $\tilde \Sigma_{k\ell}= \frac{n}{6}(1-n^2
+3m+3nm)$ with $n=t_k/\delta t$ and $m=t_\ell/\delta t$.  The matrix
$\tilde \Sigma$ is known, but the value of $c$ is unknown (it depends,
for example, on the unknown step size $\delta a$) and so $c$ must be
estimated from data.  We do this as follows.

We have $\tilde \Sigma = U \tilde Z U^T$, where
$U$ and $\tilde Z =  c^{-2} Z$ are readily computed.
Set
\begin{equation}
\tilde{r}
=
U \tilde Z^{-1/2} U^T \overrightarrow{\Delta\lambda} = c \vec{r}.
\label{eq:tilder}
\end{equation}
By the argument above, $\tilde r$
is a vector of independent Gaussians with common mean $0$
and common standard deviation $c$.  We estimate $c$
using the standard estimator
$\hat c = \sqrt{\frac{1}{K} \sum_{k=1}^K \tilde{r}_k^2}$.
The vector $\hat c^{-1} \tilde{r}$ should then be close to $\vec{r}$
when $K$ is not too small and hence should have entries that are
approximately Gaussian with common expected value $0$ and common
standard deviation $1$.  We can check this by plotting the
empirical cumulative distribution function of the entries of
$\hat c^{-1} \tilde{r}$ against the cumulative distribution
function of a Gaussian distribution with expected value $0$
and standard deviation $1$. 

In essence, we have
re-expressed the vector of measurements of an integrated Gaussian
random walk at the times $\{t_k\}$ in a new basis so that the
coefficients with respect to the new basis are independent and
identically distributed --- hence our term ``diagonalization test''.
We expect that if the $\{\Delta\lambda(t_k)\}$ arise from a two-fold
time integration of individual IID Gaussian kicks, then the entries
of $\hat c^{-1} \tilde{r} \approx \vec{r}$
will also be distributed as IID Gaussians.

Thus, the diagonalized residuals $\hat c^{-1} \tilde{r}$ provide a
convenient negative test of whether a time series is consistent with
an integrated Gaussian random walk. If the entries of the vector $\hat
c^{-1} \tilde{r}$ derived from a given observation vector
$\overrightarrow{\Delta \lambda}$ do not follow a Gaussian
distribution reasonably closely, then the time series
$\overrightarrow{\Delta \lambda}$ cannot result directly from an
integrated Gaussian random walk. Conversely, if the entries of $\hat
c^{-1} \tilde{r}$ are approximately Gaussian, then the observations
$\{\Delta \lambda(t_k)\}$ are consistent with (but do not uniquely
demand) an integrated Gaussian random walk.

The procedure outlined above is predicated on the assumption that the
longitude residuals $\{\Delta \lambda(t_k)\}$ are observed without
measurement uncertainty, so that the covariance matrix of the
observations is some multiple of $\tilde \Sigma$.  We will examine the
effects of measurement uncertainty in \S\ref{sec:test_gauss} and
\S\ref{sec:bleriot}. The diagonalization procedure also requires that
the timestep $\delta t$ be less than or equal to the time interval
between any two of the observation times $\{t_k\}$, so that the
columns of $\tilde\Sigma$ are independent.  We show in
\S\ref{sec:bleriot} that any sufficiently small $\delta t$ that
satisfies this condition will result in essentially the same
covariance matrix $\Sigma$ (and hence essentially the same probability
model for $\overrightarrow{\Delta \lambda}$), provided that $\delta a$
is modified so that the diffusivity $D = (\delta a)^2/\delta t$ is
kept constant.  Moreover, we show that any sufficiently small choice
of $\delta t$ will give essentially the same value of
$\hat{c}^{-1}\tilde{r}$ and essentially the same
diagonalization test.  However, for Bl\'eriot there is a natural,
physically motivated choice of timestep, $\delta t=\Omega^{-1}$ (see
\S\ref{sec:analytic}).

Finally, for easy reference we give a short summary of the
diagonalization test algorithm:
\begin{enumerate}
\item The input data required are a time series of values
  $\{\Delta\lambda(t_k)\}$ and the corresponding list of times
  $\{t_k\}$; the goal is to check if the $\{\Delta\lambda(t_k)\}$ are
  consistent with an integrated Gaussian random walk.
\item Assuming a constant $\delta t$ smaller than the shortest
  interval $t_{k+1} -t_k$, use $\tilde \Sigma_{k\ell}=
  \frac{n}{6}(1-n^2 +3m+3nm)$ with $n=t_k/\delta t$ and
  $m=t_\ell/\delta t$ to compute the matrix $\tilde{\Sigma}$.
\item Compute $U$ and $\tilde{Z}$ from $\tilde{\Sigma}=U\tilde{Z}U^T$
  using standard matrix decomposition procedures.
\item Compute $\tilde{r}$ from $U$, $\tilde{Z}$, and the
  $\{\Delta\lambda\}$ using Eq.~\ref{eq:tilder}.
\item Estimate $c$ using $\hat{c} =
  \sqrt{\frac{1}{K}\sum_{k=1}^{K}\tilde{r}_k^2}$, and check that
  the distribution of $\hat{c}^{-1}\tilde{r}$ has a Gaussian shape and
  a standard deviation of about 1.
\end{enumerate} 
A simple Mathematica implementation of the diagonalization test is available
in an online supplement to this paper.

\subsection{Gaussian Walks vs. L\'evy Flights} \label{sec:test_gauss}
 
As a simple check of the diagonalization test, we apply it to a
simulated integrated Gaussian random walk.\footnote{The simulations
  described here in \S\ref{sec:diagon} should not be confused with the
  3D shearing box simulations of \S\ref{sec:poisson}.}  The parameters
of our simulation are motivated by Bl\'eriot.  We take the time
interval between steps to be $\delta t = \Omega^{-1}\simeq 8842$~s and
integrate the walk for 4.3~years. Semimajor axis changes,
or ``kicks'' $\delta a$ to the moonlet, are generated as Gaussian random
variables with standard deviation 1~m.  We time-integrate $\delta a
(t)$ once to get the cumulative semimajor axis evolution $\Delta
a(t)$, and twice to get the associated longitude variations $\Delta
\lambda (t)$.  The simulated longitudes are then sampled at the
times of the Bl\'eriot observations.  When the Bl\'eriot
data are appropriately binned (see \S\ref{sec:bleriot}), there are 41
observation times.

\begin{figure*} 
\includegraphics[angle=90,width=\textwidth]{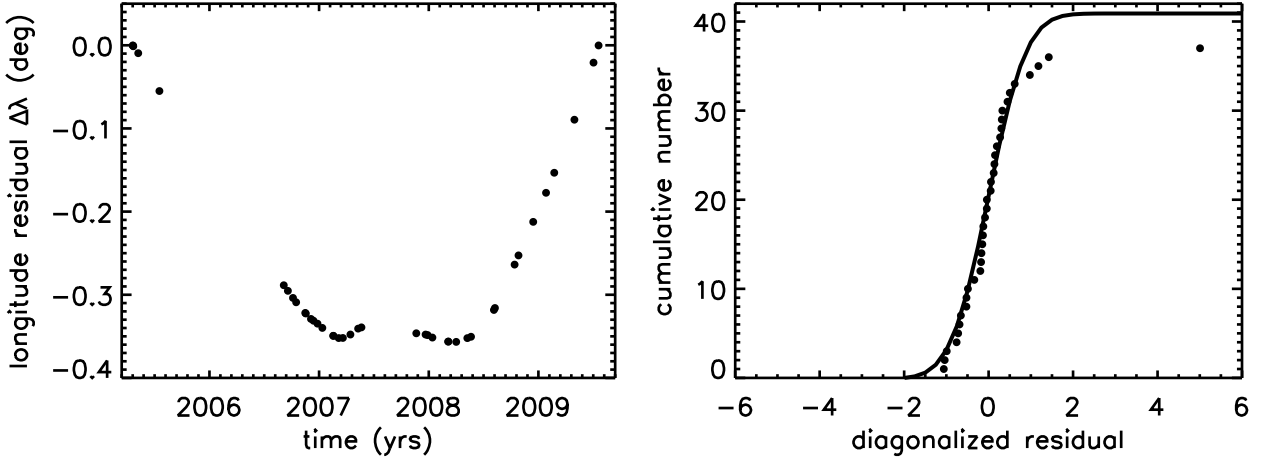} 
\caption{Checking the diagonalization test with a simulated integrated
  Gaussian random walk. A kick drawn from a Gaussian distribution of
  width 1~m in semimajor axis is applied every $\delta t =
  \Omega^{-1} = 8842$ s. The left panel shows the simulated longitude
  residuals sampled at the Bl\'eriot observation times (binned). For
  display purposes, we subtracted a linear trend to place the first
  and last data points at zero longitude residual. The right panel
  shows the results of the diagonalization test. The sorted
  diagonalized residuals (filled circles) are a reasonable match for
  Gaussian random variables (solid line). The 4 outliers dropped in
  calculating $\hat{c}$ all lie outside the range shown in this plot.
\label{fig:gaussdiag}}
\end{figure*}

Figure \ref{fig:gaussdiag} shows the results of the diagonalization
test when applied to our simulated time series containing 41
points. The diagonalized residuals $\hat{c}^{-1}\{\tilde{r}_k\}$
appear close to Gaussian.  In our calculation of $\hat{c}$ we dropped
4 extreme values among the $\{\tilde{r}_k\}$ because including these
outliers skewed $\hat{c}$ so as to be clearly inconsistent with the
vast majority of the $\{\tilde{r}_k\}$. Aside from these few outliers,
which actually fall outside the range of the right-hand panel in
Figure~\ref{fig:gaussdiag}, the agreement with a Gaussian is
satisfactory. We find $\hat{c}\simeq 1.2c$ where $c$ is computed using
the input parameters of the simulated random walk, which we also
consider satisfactory agreement.
 
\begin{figure*} 
\includegraphics[angle=90,width=\textwidth]{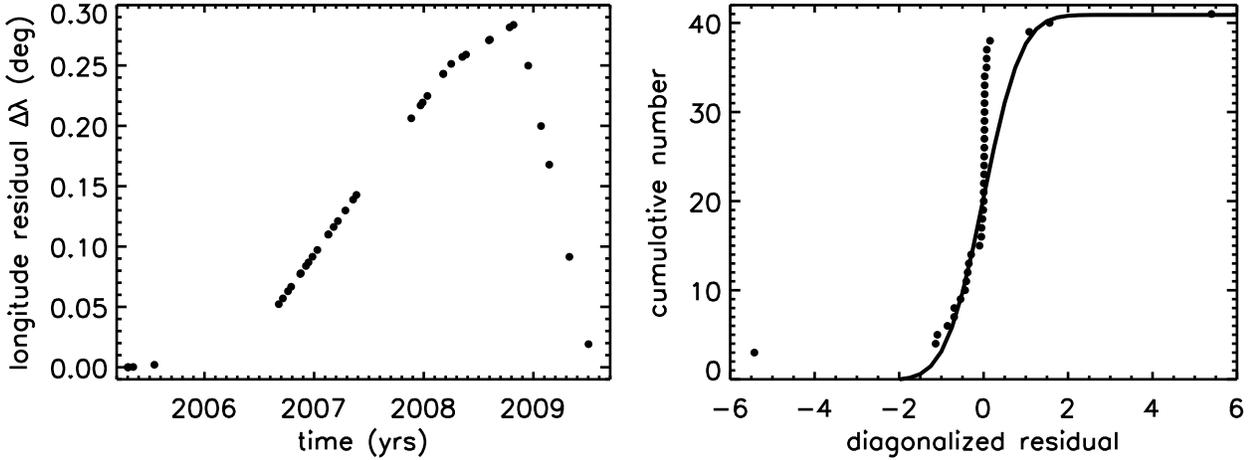} 
\caption{Checking the diagonalization test with a simulated integrated
  L\'evy flight: same as Figure~\ref{fig:gaussdiag} except that here
  the kick magnitudes $|\delta a|$ have a power-law cumulative
  distribution $\propto |\delta a|^{-2/5}$.  The sorted diagonalized
  residuals (filled circles in right panel) do not conform to a
  Gaussian distribution (solid line). In this case, 3 outliers were
  excluded in the calculation of $\hat{c}$, and all fall outside the
  range shown in the right-hand panel.
\label{fig:levydiag}}
\end{figure*}

To better calibrate what we mean by ``satisfactory,'' we also apply
the diagonalization test to data that does not derive from an
integrated Gaussian random walk. We generate instead a time-integrated
L\'evy flight, where the steps in the underlying random walk are drawn
from a power law. Specifically, we assume that the probability of
getting a kick of size $\delta a$ or larger scales as $|\delta
a|^{-2/5}$. This power law distribution describes kicks excited by
perturbers that are sparsely distributed over an annulus much wider
than the moonlet's Hill sphere radius \citep[see, for example,][and references therein]{collins06}.  Just as in the Gaussian experiment
above, we time-integrate the kicks twice to get the associated
longitude variations, and we sample the longitudes at the 41 binned
Bl\'eriot observation times. We then perform the diagonalization test
on the samples. As Figure~\ref{fig:levydiag} shows, the diagonalized
residuals are distinctly non-Gaussian in shape.
 
In a world free of measurement uncertainty, we could use $\hat{c}$ to
constrain the product of $\delta a/a$ and $\Omega \delta t$ and thus
obtain a joint constraint on the moonlet's diffusivity $D \equiv
(\delta a)^2/\delta t$.  Unfortunately, we have found by direct
experiment that $\hat{c}$ is quite sensitive to measurement
uncertainty. For example, randomly altering the longitudes shown in
the left panel of Figure \ref{fig:gaussdiag} by $\sim$2\%, an amount
comparable to the uncertainties reported for the observed Bl\'eriot
longitudes, gives $\hat{c}$ an order of magnitude larger than that
computed using the unaltered longitudes.  A similar result is obtained
for the L\'evy flight experiment in Figure
\ref{fig:levydiag}. Fortunately, the shapes of the diagonalized
residual distributions still yield the same qualitative answers: with
measurement uncertainties included, the simulated integrated Gaussian
random walk still passes the diagonalization test, and the simulated
integrated L\'evy flight still fails the test. Thus we remain
confident that, for the parameters of the problem at hand, as long as
relative measurement uncertainties remain at the level of a few
percent --- as they seem to be for the actual data of Bl\'eriot ---
the diagonalized residuals can still be used to give a ``yes-or-no''
answer to the question of whether the input data are consistent with
an integrated Gaussian random walk.

\subsection{Bl\'eriot} \label{sec:bleriot}  
 
We apply the diagonalization test to Bl\'eriot.  In mapping a given
time $t_k$ to an integer $n = t_k /\delta t$, we take the width
$\delta t$ of each time bin to equal the dynamical time $\Omega^{-1} =
8842$ s, for the physical reason that each encounter between the
moonlet and a ring particle which changes the moonlet's semimajor
axis takes a dynamical time to complete (see \S\ref{sec:analytic}).
We could take time bins of larger width, but that would reduce the
number of points in our input time series. The precise choice of
$\delta t$ is, in any case, not crucial because the random walk
is approximately scale invariant: the distribution of $\overrightarrow{\Delta \lambda}$ depends primarily on the
combination of $\delta t$ and $\delta a$ through the diffusion
coefficient $D\equiv(\delta a)^2/\delta t$. To see this, note from
Eq.~\ref{eqn:covar} that the covariance matrix $\Sigma$
of  $\overrightarrow{\Delta \lambda}$ satisfies
\[
\begin{split}
\Sigma_{k \ell}
& = 
\left(\frac{3\Omega\delta t \delta a}{2a}\right)^2 \, 
\frac{t_k}{6 \delta t}
\left(
1-
\left(\frac{t_k}{\delta t}\right)^2 
+ 3 \frac{t_\ell}{\delta t}
+ 3 \frac{t_k}{\delta t}\frac{t_\ell}{\delta t}
\right) \\
& \approx
\left(\frac{3\Omega}{2a}\right)^2 D \, 
\frac{t_k}{6}
\left(3 t_k t_\ell - t_k^2\right) \\
\end{split}
\]
for $t_k \le t_\ell$. Hence, the probability model for $
\overrightarrow{\Delta \lambda}$ is, to first order, unaffected by a
change in $\delta t$, provided that $\delta t$ is sufficiently small
and $\delta a$ is also changed so as to keep the diffusivity $D$
constant.  The same calculation shows that if $\delta t$ is changed to
$h \delta t$ for some constant $h$, then $\tilde \Sigma$ is, to first
order, changed to $h^{-3} \tilde \Sigma$.  Then $\tilde r$ and $\hat
c$ become respectively $h^{3/2} \tilde r$ and $h^{3/2} \hat c$, so
that $\hat c^{-1} \tilde r$ is, to first order, unchanged.
Consequently, the diagonalization test is effectively invariant with
respect to a change in the choice of $\delta t$.

\begin{figure*}  
\begin{center} 
\includegraphics[angle=90,width=\textwidth]{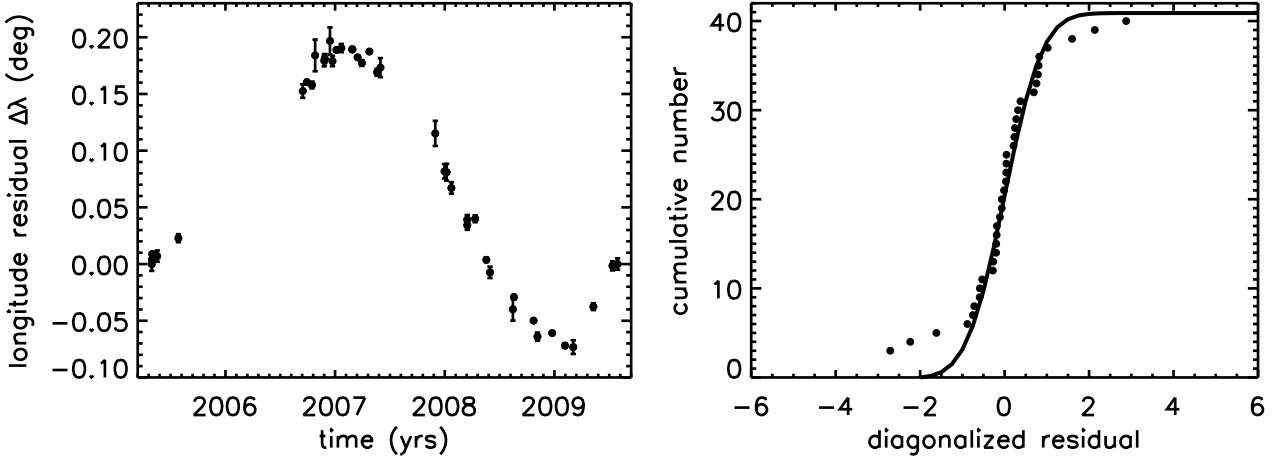}  
\end{center}  
\caption{Results of the diagonalization test for Bl\'eriot: same as
  Figure~\ref{fig:gaussdiag} but for the actual data for Bl\'eriot
  rather than for simulation results. The data are binned into 41
  points such that the time interval between bins is at least
  $\Omega^{-1} = 8842$ s long. The diagonalized residuals are
  reasonably close to Gaussian-distributed; compare with
  Figures~\ref{fig:gaussdiag} and \ref{fig:levydiag}. As in previous figures,
the 3 outliers excluded from the calculation of $\hat{c}$ fall 
outside the range shown in the right panel.}
\label{fig:bleriotdiag}  
\end{figure*}

Many points in Bl\'eriot's published time series are separated by less
than $\delta t = \Omega^{-1}$, as multiple images were taken in short
succession.  The longitudes that fall into the same time bin are
averaged into one number. Measurement uncertainties are added in
quadrature.  Binned this way, there are 41 ``independent'' longitude
measurements, as shown in Figure \ref{fig:bleriotdiag} (left
panel).\footnote{Within some bins, the original longitudes differ by
  more than their published error bars (see, e.g., the cluster of five
  points near year 2005.38 in Figure 4b of T10). The published
  uncertainties are probably underestimated, especially in those cases
  where longitudes are referenced to ring features such as the Encke
  gap edge instead of to stars.  We have verified with M.~Tiscareno
  (2012, personal communication) that our binning procedure is
  justified given the quality of the data. \label{foot:matt}}

The matrix operations in Eq.~\ref{eqn:decorr} are applied to the
binned longitude residuals $\{ \Delta\lambda(t_k) \}$ to compute the
diagonalized residuals $\hat{c}^{-1}\{\tilde{r}_k \}$. 
In this case, 3 outlier
values were dropped in the computation of $\hat{c}$. Figure
\ref{fig:bleriotdiag} shows the distribution of diagonalized
residuals. It is close to Gaussian; compare with Figures
\ref{fig:gaussdiag} and \ref{fig:levydiag}.

We have explored the sensitivity of these results 
to measurement uncertainties in Bl\'eriot's data.
Ten different realizations of Bl\'eriot's longitude
time series were generated by randomly selecting points within
the error bars shown in the left panel of Figure \ref{fig:bleriotdiag}.
In all cases, the diagonalized residuals resembled those
shown in the right panel of Figure \ref{fig:bleriotdiag}.

Recall our experiments in \S\ref{sec:test_gauss} with simulated random
walks, where we found that the magnitude of the diagonalized residuals
was sensitive to measurement uncertainty; in particular, the
(unnormalized) diagonalized residuals were much larger than expected
when measurement errors were added to the random walks. Indeed the
same effect seems to manifest here with the actual data for Bl\'eriot.  
The $\hat{c}$ value for Bl\'eriot is an order of magnitude larger than
the $\hat{c}$ value for the simulated integrated Gaussian walk even
though their longitude residuals are of comparable magnitudes.
Thus we have no reliable estimator of the true value of $c$, and 
thus no constraint on the moonlet's diffusivity $D$ from the results of
the diagonalization test alone.

Although we cannot measure the moonlet's diffusivity 
from the diagonalization test,
we still have the original
longitude time series of Bl\'eriot, in addition to a smattering of
longitude data for other propellers (see Table 1 of T10). Taken at
face value, these data indicate that moonlet diffusivities must be
such as to generate ``typical'' longitude deviations of $\sim$0.1--0.3
deg over timescales of $\sim$1--2 yr.  In the next section, we explain
how such random walks can physically arise.
 
\section{Stochastic Migration due to Local Fluctuations in Surface Density}\label{sec:poisson}  
  
A moonlet can undergo a Gaussian random walk because of Gaussian
fluctuations in the number of particles encountered at Hill sphere
separations.  The fluctuations of interest to us do not depend on ring
self-gravity. Fluctuations still occur because of frequent collisions
between ring particles that randomize their positions in the time intervals
between encounters with the moonlet.  We sketch this process analytically
(\S\ref{sec:analytic}), test and calibrate our order-of-magnitude
scaling relations with numerical $N$-body simulations
(\S\ref{sec:numerical}), and apply our theory of stochastic migration
to Bl\'eriot and other propellers (\S\ref{sec:bleriot_conclude}). Many
of the ideas in this section have been treated previously (e.g.,
\citealt{murrayclay06}; \citealt{rein10}; \citealt{crida10}), but we
present them here afresh for clarity and convenience.
 
\subsection{Analytic Description of Gaussian Stochastic Migration}\label{sec:analytic}  
  
Consider a moonlet of radius $R_{\rm moon}$ at semimajor axis $a$,
embedded in a ring composed of particles each of radius $r$ and mass
$m$. 
The moonlet and ring particles are assumed individually to have a bulk
density $\rho_b$.  The surface mass density of the ring is
$\Sigma$, and the local orbital frequency is $\Omega$. Ring particles
shear by the moonlet and gravitationally perturb it. Particles inside
the moonlet's orbit tend to kick the moonlet onto a larger orbit,
while particles outside the moonlet's orbit tend to push the moonlet
inward.
 
Random fluctuations in the rate of particles encountered cause the 
moonlet's semimajor axis to change stochastically. At radial 
separations on the order of $x$ between a collection of ring particles 
and the moonlet, the relative Keplerian shearing velocity is $\sim 
\Omega x$.  The duration of an encounter is $\delta t_{\rm enc} \sim x 
/ (\Omega x) \sim \Omega^{-1}$, independent of $x$.  The number of 
particles passing conjunction (to either side of the moonlet's orbit) 
per $\delta t_{\rm enc}$ should follow a Poisson distribution with 
mean $N_{\rm enc} \sim \Sigma x^2 / m$ and width 
$\sqrt{N_\mathrm{enc}}$. 
 
The randomly varying excess number of particles --- lying to either
side with equal probability --- is responsible for net changes in the
moonlet's semimajor axis.  We call the mass of this excess group of
particles, tallied every encounter time $\delta t_{\rm enc}$, the
``fluctuation mass'' $m_{\rm fluct}$. If $N_\mathrm{enc}\gg 1$, we can
treat $m_{\rm fluct}$ as approximately Gaussian-distributed with mean
zero and width $\sim$$m \sqrt{N_{\rm enc}}$: positive/negative signs
correspond to excess groups of particles passing outside/inside the
moonlet's orbit. An encounter with a fluctuation mass occurs once
every $\delta t_{\rm enc}$, and each encounter changes the moonlet's
velocity by $\delta v \sim (G m_{\rm fluct} / x^2) \times \delta
t_{\rm enc}$, with $G$ the gravitational constant.  The largest
fluctuations arise from particles within several Hill sphere radii of
the moonlet \citep{murrayclay06,crida10}: $x \sim R_{\rm Hill} =
a\left(\frac{4\pi}{9}\rho_b
R_\mathrm{moon}^3/m_\mathrm{Saturn}\right)^{1/3}$.  For such
encounters, the fractional change in the moonlet's semimajor axis is
of order the fractional change in its velocity: $\delta a/a \sim
\delta v / (\Omega a)$, with equal probability of either
sign.\footnote{This relation holds for $x \sim R_{\rm Hill}$, but not
  for $x \gg R_{\rm Hill}$.}
 
Putting all of the above together, we find that every 
$\delta t_{\rm enc} \sim \Omega^{-1}$ time, the moonlet 
randomly walks in semimajor axis by a step of root-mean-square size 
\begin{align}  
\delta a & \sim \frac{m_{\rm fluct}}{m_{\rm Saturn}} \left( \frac{a}{R_{\rm Hill}} \right)^2 a \nonumber \\  
 & \sim 0.15 \left( \frac{300 \, {\rm m}}{R_{\rm moon}} \right) \left( \frac{r}{10 \, {\rm m}} \right)^{3/2} \nonumber\\
 &\quad\quad\quad\quad\cdot\left( \frac{\Sigma}{40 \g/{\rm cm}^2} \right)^{1/2} \left( \frac{1 \g/{\rm cm}^3}{\rho_b} \right)^{1/6} \, {\rm m}  \,.\label{eqn:step_gauss}   
\end{align}  
Over the course of the {\it Cassini} observations analyzed by T10, a
propeller-moonlet will random walk in semimajor axis by an rms
distance
\begin{equation}  
\Delta a (t) \sim \delta a \sqrt{\Omega t} \sim 10 \left( \frac{300 \m}{R_{\rm moon}} \right) \left( \frac{r}{10 \, {\rm m}} \right)^{3/2} \left( \frac{t}{2 \yr} \right)^{1/2} \m \,  
\label{eqn:walk_gauss}  
\end{equation}  
where we have used $\Sigma=40\;\g\;\cm^{-2}$ and
$\rho_b=1\;\g\;\cm^{-3}$.  Comparison with Eq.~\ref{eqn:delta_a} shows
that this is of the right order of magnitude to explain the observed
non-Keplerian motions of propellers.  In the next section we employ
$N$-body simulations to test our scaling relations and measure more
accurately the moonlet's diffusivity. That is, we will revise our
crudely estimated coefficients in Eqs.~\ref{eqn:step_gauss} and
\ref{eqn:walk_gauss} in accord with the numerical simulations.
   
\subsection{Numerical Simulations of Gaussian Stochastic Migration}\label{sec:numerical}  
\begin{figure}  
\begin{center} 
\includegraphics[width=\columnwidth]{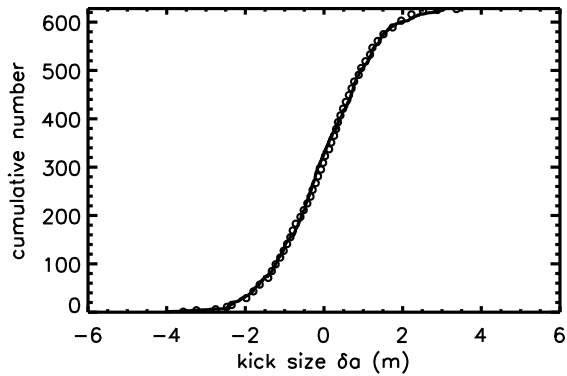} 
\end{center}  
\caption{Distribution of kicks (changes in semimajor axis) felt by
  our simulated 100-m moonlet embedded in a ring with surface density
  $\Sigma = 40$~g cm$^{-2}$, maximum particle size $r_{\rm max} =
  10$~m, minimum particle size $r_{\rm min} = 2.5$~m, and power-law
  index $q=3$ for the differential size distribution of particles.
  Kicks are computed every time interval $\Omega^{-1}$; their
  cumulative distribution (solid line) conforms closely to a Gaussian
  (open circles).  The typical kick size, $\delta a
  \sim 1.5$~m, is consistent with our order-of-magnitude estimate in
  Eq.~\ref{eqn:step_gauss}.}
\label{fig:hannokicks}  
\end{figure}  
 
We perform shearing box simulations of a moonlet randomly perturbed by 
nearby ring particles.  We use the freely available collisional 
$N$-body code \texttt{REBOUND} \citep{rein12} with shear periodic 
boundary conditions and the symplectic epicycle integrator 
\citep[SEI,][]{rein11}. 
 
The simulations are similar to the test-particle simulations performed 
by \cite{crida10} but include particle-particle and particle-moonlet 
collisions.  Unlike the simulations of \cite{rein10} and 
\cite{lewis09}, ours do not explicitly include self-gravity. 
As explained in \S\ref{sec:intro}, self-gravity wakes probably make only a 
small contribution to the observed longitude residuals of propellers 
as large as Bl\'eriot. The mean self-gravitational field 
does enhance the vertical frequency $\Omega_z$ of epicyclic motion (e.g., \citealt{wisdom88}); 
we mock up this effect in the code by setting $\Omega_z = 3.6 \Omega$. 
All simulations are performed with $\Omega = 1.131\cdot 10^{-4} 
\s^{-1}$ corresponding to a semimajor axis of 
$a=130000\,\mathrm{km}$.  The timestep was chosen to be 
$dt=10^{-3}\,2\pi/\Omega$. 
 
Simulation parameters are listed in Table~\ref{tab:nbody}.  
The ring particles and moonlet are assumed to have a bulk density 
$\rho_b = 0.4$ g/cm$^3$.  Ring particles are assumed to follow a 
differential size distribution $dN/dr \propto r^{-q}$ from 
$r_\mathrm{min}$ to $r_\mathrm{max}$.    
The slope $q$ is fixed at 3, which places most of the mass 
in the largest ring particles. Some of our chosen particle size 
parameters are compatible with occultation and 
imaging observations (Cuzzi et al.~2009; T10). 
Others were chosen only to provide a large enough dynamic range to probe 
how stochasticity scales with the particle radius~$r$. 
 
Ring particles are initialized with zero random velocity (i.e., their
initial velocity is determined purely by Keplerian shear).  Once a
ring particle exits an azimuthal boundary of the simulation domain, it
re-enters the domain on the opposite side at a randomized radial
location (semimajor axis), with zero random velocity.  Thus the number
of ring particles $N$ in the box remains constant.  The dimensions of the
box are $L_x$, $L_y$, and $L_z$ in the radial, azimuthal, and vertical
directions, respectively.  In nearly all cases, $L_x \times L_y$
covers $\sim$$27 \times 135$ moonlet Hill radii, while $L_z$ is chosen
large enough so that no particle ever reaches a vertical
boundary.
 
\begin{table*} 
\centering 
\begin{tabular}{c | cccccc | cccccccccc} 
Sim. Number  & $R_\mathrm{moon}$ & 	$\Sigma$ &	$\rho_b$ &  	$r_{\mathrm{min}}$ &	$r_{\mathrm{max}}$ &	$q$ & $L_x$ &	$L_y$ &	$L_z$ & $N$ & Time \\ 
 & (m) & (g/cm$^2$) & (g/cm$^3$) & (m) & (m) &  & (m) & (m) & (m) & & $(1/\Omega)$ \\ \hline\hline 
1 	& $100$ 	& $40$ 	& $0.4  $	& $2.5$    &  $95  $ & 3 & $3500$ & $17500$ & 1000 & 19504 & 540.6 \\  
2 	& $100$ 	& $40$ 	& $0.4  $	& $2.5$    &  $85  $ & 3 & $3500$ & $17500$ & 1000 & 21068 & 540.6 \\  
3 	& $100$ 	& $40$ 	& $0.4  $	& $2.5$    &  $75  $ & 3 & $3500$ & $17500$ & 1000 & 21992 & 540.6 \\  
4 	& $100$ 	& $40$ 	& $0.4  $	& $2.5  $    &  $55  $ & 3 & $3500$ & $17500$ & 1000 & 23699 & 540.6 \\  
5 	& $100$ 	& $40$ 	& $0.4  $	& $2.5  $    &  $45  $ & 3 & $3500$ & $17500$ & 1000 & 27968 & 540.6 \\  
6 	& $100$ 	& $40$ 	& $0.4  $	& $2.5  $    &  $35  $ & 3 & $3500$ & $17500$ & 1000 & 36662 & 540.6 \\  
7 	& $100$ 	& $40$ 	& $0.4  $	& $2.5  $    &  $25  $ & 3 & $3500$ & $17500$ & 1000 & 51925 & 540.6 \\  
8 	& $100$ 	& $40$ 	& $0.4  $	& $2.5  $    &  $15  $ & 3 & $3500$ & $17500$ & 1000 & 90656 & 540.6 \\  
9 	& $100$ 	& $40$ 	& $0.4  $	& $2.5  $    &  $10  $ & 3 & $3500$ & $17500$ & 1000 & 145912 & 540.6 \\  
10 	& $100$ 	& $40$ 	& $0.4  $	& $2.5  $    &  $5  $ & 3 & $3500$ & $17500 $ & 1000 & 350177 & 540.6 \\  
11 	& $100$ 	& $40$ 	& $0.4  $	& $2.5  $    &  $2.5  $ & 3 & $3500$ & $17500 $ & 1000 & 935860 & 540.6 \\  
12  	& 50 & 40 & 0.4 & 14.5 & 15 & 3 & 1750 & 8750 & 1000 & 1121 & 540.6 \\ 
13  	& 100 & 40 & 0.4 & 14.5 & 15 & 3 & 3500 & 17500 & 1000 & 4390 & 522.6 \\ 
14  	& 200 & 40 & 0.4 & 14.5 & 15 & 3 & 7000 & 35000 & 1000 & 17416 & 540.6 \\ 
15  	& 400 & 40 & 0.4 & 14.5 & 15 & 3 & 14000 & 70000 & 1000 & 69545 & 540.6 \\ 
16  	& 800 & 40 & 0.4 & 14.5 & 15 & 3 & 28000 & 140000 & 1000 & 277980 & 473.6 \\ 
17  	& 100 & 40 & 0.4 &  2.5 & 10 & 3 &  1000 &   2000 & 1000 &   4812 & 20738 \\ 
\end{tabular} 
\caption{Parameters of simulations using \texttt{REBOUND}, a shearing box code for colliding particles. The duration of each simulation is given in units of $1/\Omega$. See text for description. \label{tab:nbody}} 
\end{table*} 
 
Figure~\ref{fig:hannokicks} shows the distribution of semimajor axis
changes or ``kicks'' $\delta a$, evaluated every time interval $\delta
t= 1/\Omega$, for simulation 9. The empirical distribution of kicks is
close to Gaussian, confirming our physical description of stochastic
migration in \S\ref{sec:analytic}. Furthermore, the characteristic
value of $\delta a$ (defined as the $1\sigma$ half-width of the
Gaussian distribution) is $1.5$~m, which agrees to order-of-magnitude
with the prediction of Eq.~\ref{eqn:step_gauss} for a 100-m
moonlet. In fact, the simulated characteristic value for $\delta a$ is
about 3 times larger than predicted by our back-of-the-envelope
estimate. The enhanced fluctuations revealed by the numerical
simulations strengthen the case for perturbations from the largest
(decameter-sized) ring particles (\S\ref{sec:bleriot_conclude}) as the
main cause of the propellers' observed non-Keplerian motions. We
speculate that the factor of 3 might arise from two effects in our
numerical simulations that are omitted from our simple analysis in
\S\ref{sec:analytic}: ring particles on horseshoe orbits, and direct
collisions between ring particles and the moonlet (\citealt{rein10};
see also \citealt{murrayclay06} who show that encounters with
horseshoe librators enhance stochasticity by a factor of order unity).

\begin{figure*}  
\begin{center} 
\includegraphics[angle=90,width=\textwidth]{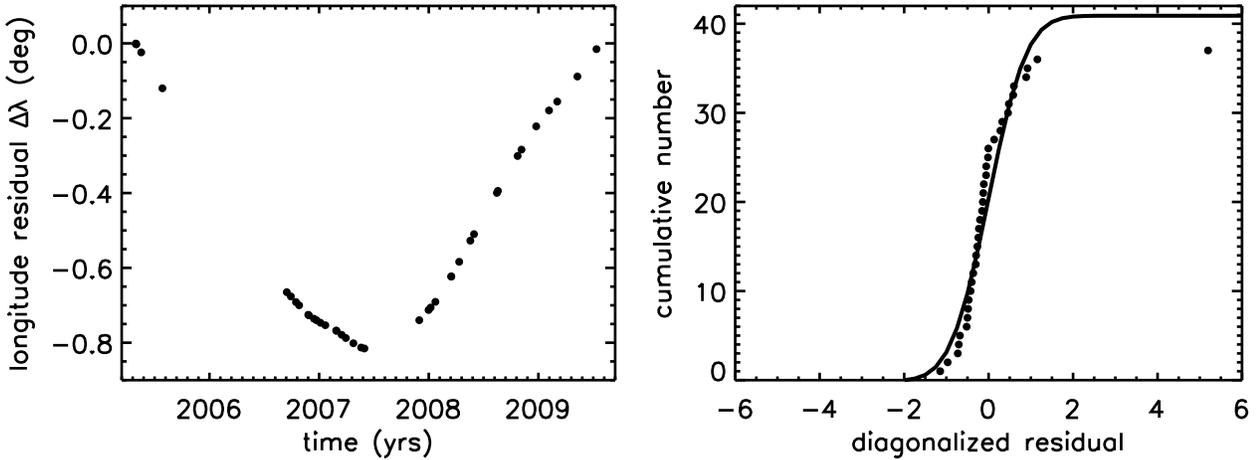} 
\end{center}  
\caption{Results of the diagonalization test for simulation 17: same
  as Figure~\ref{fig:gaussdiag} but for simulation 17 sampled at the
  41 binned observation times of Bl\'{e}riot. Here 4 outliers were
  excluded from the calculation of $\hat{c}$, and all lie outside the
  range shown in the right-hand panel. The longitude residuals shown
  in the left panel are simulation data sampled at times corresponding
  to the Bl\'eriot observation times. The residuals' distribution is
  close to Gaussian.
}
\label{fig:simdiag}  
\end{figure*}

Figure~\ref{fig:simdiag} shows how the moonlet's longitude residuals
evolve in simulation 17, whose parameters are the same as those of
simulation 9 but is run for 5 years. In order to avoid impractically
long runtimes, simulation 17 uses a smaller box size than simulation
9. However, we expect the results to be insensitive to the
change in box size: the most important interactions are with
particles passing within $\sim$$R_\mathrm{Hill}$ of the moonlet
(\S\ref{sec:analytic}) and the smaller box is still several
$R_\mathrm{Hill}$ in size. We applied the diagonalization test to
simulation 17, sampling the longitude residuals at the same set of 41
times that characterize Bl\'eriot's binned data. The diagonalized
residuals, shown in Figure~\ref{fig:simdiag}, verify that the behavior
of the simulated moonlet is consistent with that of an integrated
Gaussian random walk.

\begin{figure*} 
\begin{center} 
\includegraphics[width=\columnwidth]{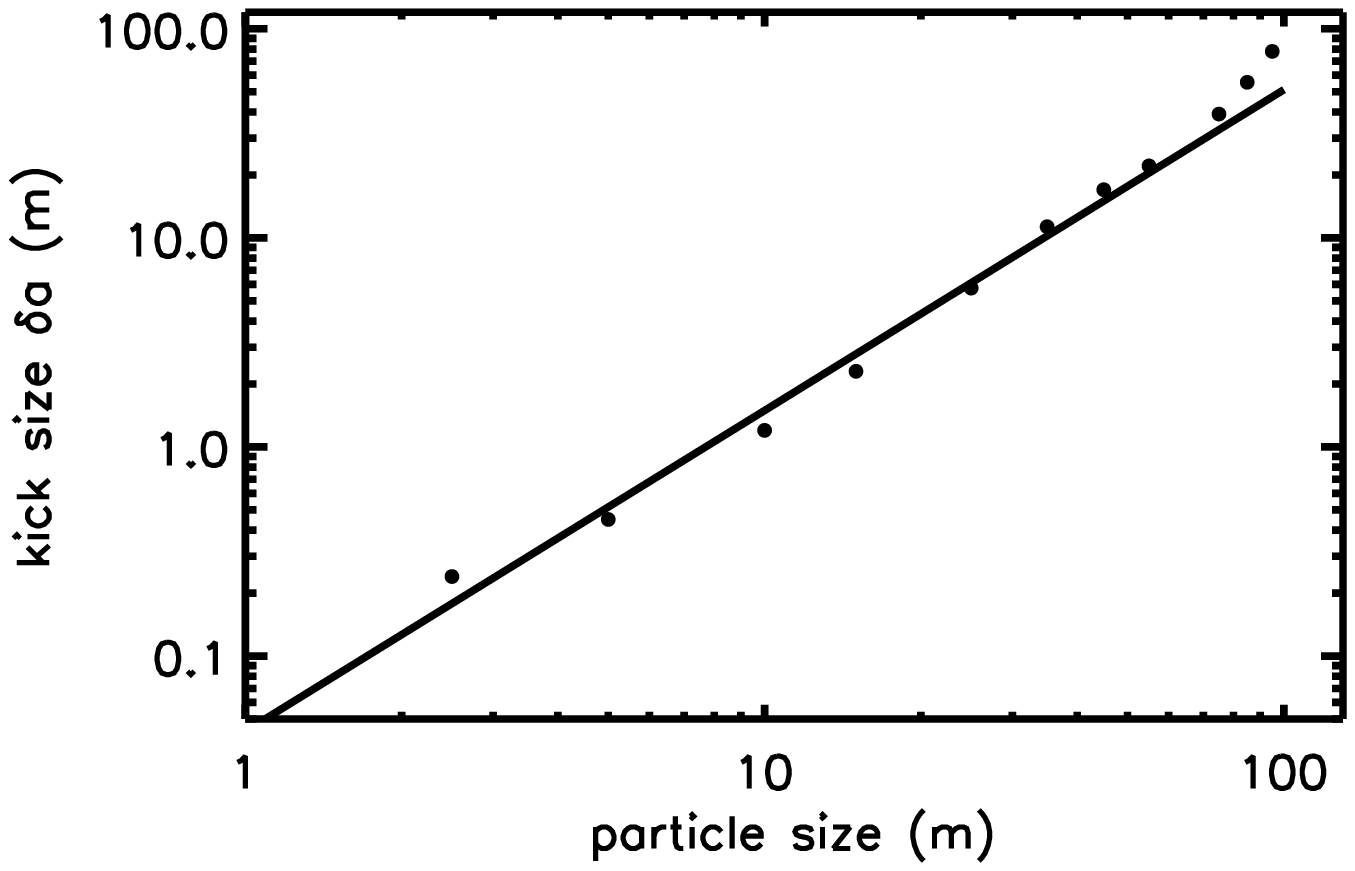} 
\includegraphics[width=\columnwidth]{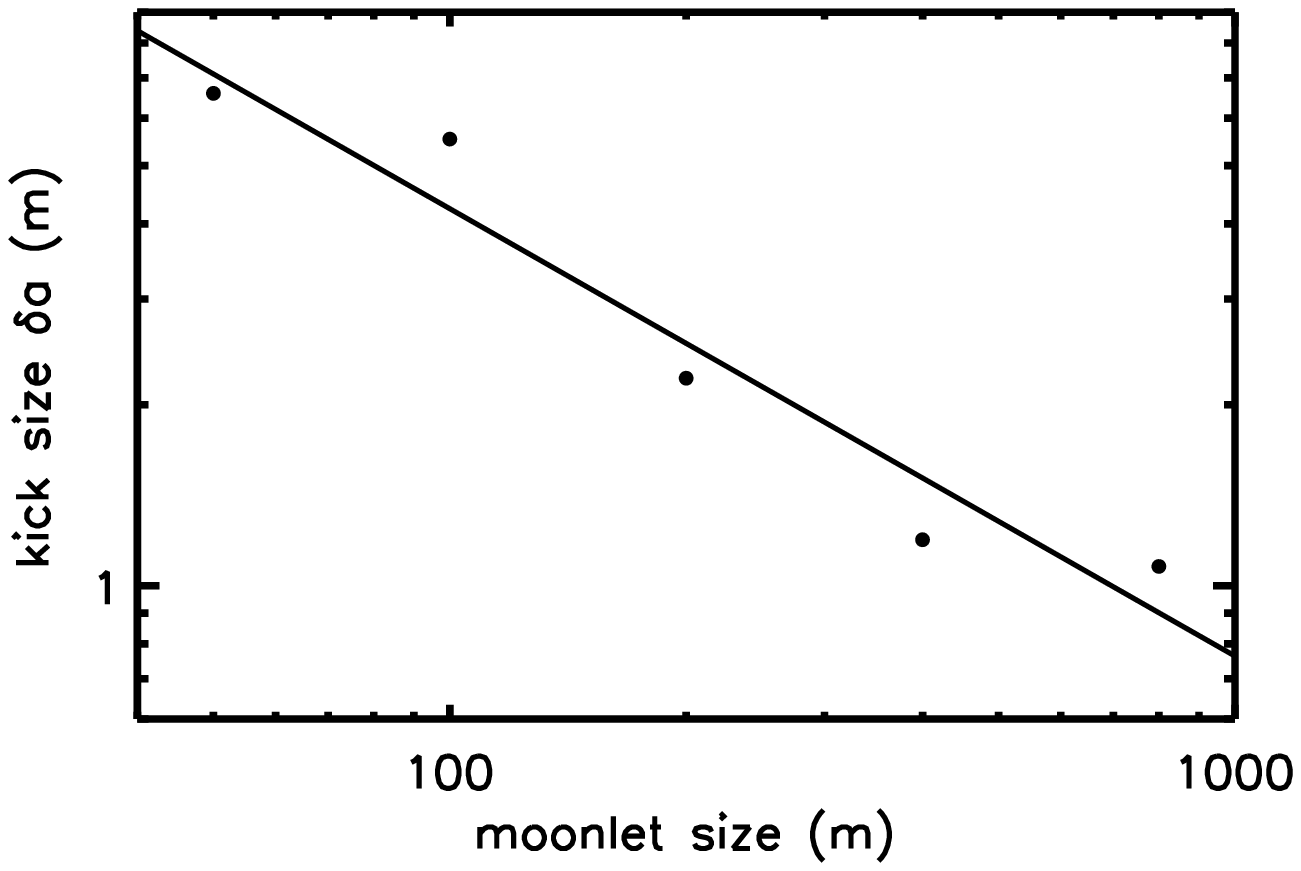} 
\end{center} 
\caption{Characteristic $\delta a$ value, or ``kick size'', for
  simulations 1--11 (left panel) and 12--16 (right panel). Simulations
  1--11 show how varying the maximum particle size affects the typical
  $\delta a$. The best-fit power law shown has index 1.47.  This
  agrees well with the predicted slope of 1.5 from Eq.~\ref{eqn:step_gauss}. The
  fit uses only data from simulations 6--11: in simulations 1--5, the
  largest ring particles are so rare that $N_\mathrm{enc}\lesssim
  1$. Simulations 12--16 show how varying the moonlet size affects the
  typical $\delta a$. The best-fit power law shown has slope $-0.74$;
  given the scatter in the data, we consider this acceptable agreement
  with Eq.~\ref{eqn:step_gauss}, which predicts a slope of $-1$.
  \label{fig:rtrends}}
\end{figure*}

Figure~\ref{fig:rtrends} shows the characteristic value of $\delta a$
versus both particle radius $r$ and moonlet radius $R_{\rm moon}$.  We
compute this characteristic value by plotting a cumulative Gaussian
distribution against the kicks sorted from smallest to largest,
performing a least-squares fit to a line, and varying the width of the
Gaussian until the slope of the fitted line is 1. We take the Gaussian
width so derived to be the characteristic $\delta a$. The scalings
expected from Eq.~\ref{eqn:step_gauss} are approximately
confirmed. The agreement for $r$ is better than for $R_{\rm moon}$,
but we consider both acceptable. Using our numerical simulations to
normalize our analytic scalings, we calibrate
Eqs.~\ref{eqn:step_gauss} and \ref{eqn:walk_gauss} into more accurate
forms:
\begin{align}
\delta a &\approx 0.5 \left( \frac{300 \, {\rm m}}{R_{\rm moon}} \right) \left( \frac{r}{10 \, {\rm m}} \right)^{3/2}\nonumber\\
&\quad\quad\quad\cdot \left( \frac{\Sigma}{40 \g/{\rm cm}^2} \right)^{1/2} \left( \frac{1 \g/{\rm cm}^3}{\rho_b} \right)^{1/6} \, {\rm m}  \label{eqn:step_gauss_calib}   
\end{align}
and
\begin{equation}  
\Delta a (t) \sim \delta a \sqrt{\Omega t} \approx 30 \left( \frac{300 \m}{R_{\rm moon}} \right) \left( \frac{r}{10 \, {\rm m}} \right)^{3/2} \left( \frac{t}{2 \yr} \right)^{1/2} \m \,.  
\label{eqn:walk_gauss_calib}  
\end{equation}  
 
\subsection{Implications for Bl\'eriot}\label{sec:bleriot_conclude}  
 
Observations of Bl\'eriot require $\Delta a \simeq 30$ m
over a period of $\sim$2 years. We can use 
Eq.~\ref{eqn:walk_gauss_calib}, which is calibrated
using numerical simulations, to estimate how big the
surrounding ring particles must be to reproduce these observed
parameters.

The radius of Bl\'eriot's moonlet is thought to lie in the range
$R_{\rm moon} = 300$--1200 m (see Figure 2 of T10).  If we adopt
$R_{\rm moon} = 700$ m, then a particle size of $r \simeq 18$ m would
satisfy the observations assuming a bulk density of 1 g/cm$^3$ for
all bodies.  That is, the 1-$\sigma$ excursion in semimajor axis for
a 700-m moonlet is $\Delta a\simeq 30$~m over $\Delta t = 2$ yr 
when $r = 18$~m.  If the
observed $\Delta a \simeq 30$~m actually represents a 2-$\sigma$
excursion, then the required particle size decreases to $r = 11$~m.

In fact, ground- and space-based occultation data of the outer A ring
independently indicate that the bulk of the ring mass resides in
particles of size $r=10$--20~m \citep[for a summary of what is known
about particle size distributions based on occultation analysis,
see][]{cuzzi09}.  For the region just outside the Encke gap which
contains Bl\'eriot and the other giant propellers, fits to Voyager
observations yield a maximum particle size $r_{\rm max} = 8.9$~m and
$q=3.03$ \citep{zebker85}. An analysis of ground-based occultation
data gives $r_{\rm max} = 20$~m and $q=2.9$ \citep{french00}. We
conclude that stochastic gravitational interactions between propeller
moonlets and the largest nearby particles in the outer A ring can
readily reproduce longitude residuals like those observed for
Bl\'eriot.
 
\section{SUMMARY}  
\label{sec:sum}   

Whether the migration patterns of propellers arise from a
deterministic or random process is not obvious just by looking at their
longitude residuals.  The difficulty arises because 
longitude residuals are time-integrated quantities. The time
integration smooths out semimajor axis variations that could be
noisy, and introduces correlations between data at a given time
and all earlier times.

The ``diagonalization test'' removes correlations introduced
by time integration of a Gaussian random process. It tests whether
a given time series is compatible with an integrated Gaussian random
walk. We have applied the diagonalization test to the longitude
time series of the propeller Bl\'eriot and found that it passes
the test. Bl\'eriot's behavior is consistent with that of an integrated
Gaussian random walk.

By combining simple analytic scaling relations with numerical $N$-body
simulations, we also showed that moonlets as large as Bl\'eriot,
having radii of $\sim$700 m, could exhibit longitude residuals on the
order of 0.1 deg over 2 years, when embedded in a ring of surface
density $40$ g/cm$^2$---provided the bulk of the mass of the ring is
contained in particles 10--20 m in radius. Such ring properties are
inferred on independent grounds by occultation analysis \citep{cuzzi09}. The perturbations exerted by large ring particles on
propeller-moonlets are stochastic, caused by Poisson fluctuations in
the number of ring particles that shear by the moonlet on Hill sphere
scales. 

The picture of stochastic migration that we support is
similar to that first proposed by \cite{rein10}, except
that the primary contributors to stochasticity are decameter-sized
particles, not self-gravity wakes. We have shown by direct $N$-body
simulation (e.g., Figure \ref{fig:simdiag}) that particle size
distributions that place most of the ring mass in decameter sizes can
reproduce longitude residuals like those observed. As the {\it
  Cassini} spacecraft emerges from the ring plane in 2012 and resumes
observations of propellers, we look forward to measurements of
longitude time series for other propellers in addition to
Bl\'eriot---and to more accurate protocols for making longitude
measurements by improvements to the matrix describing the orientation
of the ISS camera with respect to the spacecraft.  These new and more
accurate data can also be subjected to the diagonalization test, and
used to test our prediction that longitude residuals scale with
moonlet size as $\Delta \lambda \propto R_{\rm moon}^{-1}$.
 
\section*{\small{Acknowledgments}}
\small{ We are grateful to Matt Tiscareno for illuminating
  conversations about the data published in T10; Scott Tremaine for
  helpful comments at various stages of the project; Philip Stark for
  arranging this cross-talk between astronomy and statistics; and an
  anonymous referee for a thoughtful and constructive report. MP and
  EC acknowledge support from NSF grant AST-0909210, NASA Outer
  Planets Research grant NNX12AJ09G, Berkeley's Center for Integrative
  Planetary Science, and Berkeley's Theoretical Astrophysics Center.
  HR was supported by the Institute for Advanced Study and NSF grant
  AST-0807444, and SNE acknowledges support from NSF grant
  DMS-0907630. Some of this work was begun as part of the
  International Summer Institute for Modeling in Astrophysics held at
  the Kavli Institute for Astronomy and Astrophysics in Beijing
  University. We thank the organizers of this summer program,
  including Pascale Garaud, Doug Lin, and Shang-Fei Liu for their
  tireless efforts at fostering collaborations, and are indebted to
  participants Peng Jiang and Zhao Sun for enlightening conversations.
}

\bibliographystyle{aa}  
\bibliography{tadpole}  
  
\end{document}